\documentclass[sigconf]{acmart}

\renewcommand\footnotetextcopyrightpermission[1]{}

\setcopyright{acmlicensed}
\copyrightyear{2024}
\acmYear{2024}
\acmDOI{XXXXXXX.XXXXXXX}

\acmConference[ASE '24 - To appear]{39th IEEE/ACM International Conference on Automated Software Engineering}{October 27--November 1, 2024}{Sacramento, CA, USA}

\acmISBN{978-1-4503-XXXX-X/18/06}

\usepackage{amsmath,amsfonts}
\usepackage{algorithmic}
\usepackage{graphicx}

\usepackage{draftwatermark}
\SetWatermarkText{Preprint}
\SetWatermarkFontSize{4cm}

\usepackage{macros}

\begin{document}

\title{DeepREST: Automated Test Case Generation for REST APIs Exploiting Deep Reinforcement Learning}

\author{Davide Corradini}
\email{davide.corradini@univr.it}
\orcid{0009-0009-2594-4562}
\affiliation{
  \institution{University of Verona - Dept. of Computer Science}
  \city{Verona}
  \country{Italy}
}

\author{Zeno Montolli}
\email{zeno.montolli@univr.it}
\affiliation{
  \institution{University of Verona - Dept. of Computer Science}
  \city{Verona}
  \country{Italy}
}

\author{Michele Pasqua}
\email{michele.pasqua@univr.it}
\orcid{0000-0002-9475-4836}
\affiliation{
  \institution{University of Verona - Dept. of Computer Science}
  \city{Verona}
  \country{Italy}
}

\author{Mariano Ceccato}
\email{mariano.ceccato@univr.it}
\orcid{0000-0001-7325-0316}
\affiliation{
  \institution{University of Verona - Dept. of Computer Science}
  \city{Verona}
  \country{Italy}
}

\begin{abstract}
    Automatically crafting test scenarios for REST APIs helps deliver more reliable and trustworthy web-oriented systems. However, current black-box testing approaches rely heavily on the information available in the API's formal documentation, i.e., the OpenAPI Specification (OAS for short).  %
    While useful, the OAS mostly covers syntactic aspects of the API (e.g., producer-consumer relations between operations, input value properties, and additional constraints in natural language), and it lacks a deeper understanding of the API business logic. Missing semantics include implicit ordering (logic dependency) between operations and implicit input-value constraints. These limitations hinder the ability of black-box testing tools to generate truly effective test cases automatically.

    This paper introduces \deeprtg, a novel black-box approach for automatically testing REST APIs. It leverages \emph{deep reinforcement learning} to uncover implicit API constraints, that is, constraints hidden from API documentation. %
    Curiosity-driven learning guides an agent in the \emph{exploration} of the API and learns an effective order to test its operations. This helps identify which operations to test first to take the API in a testable state and avoid failing API interactions later. At the same time, \emph{experience} gained on successful API interactions is leveraged to drive accurate input data generation (i.e., what parameters to use and how to pick their values). Additionally, \deeprtg alternates exploration with \emph{exploitation} by mutating successful API interactions to improve test coverage and collect further experience.

    Our empirical validation suggests that the proposed approach is very effective in achieving high test coverage and fault detection and superior to a state-of-the-art baseline.
\end{abstract}



\maketitle

\begin{center}
	\begin{tikzpicture}
		\node[anchor=west,text width=.46\textwidth] (A) at (0,0) {Paper accepted for publication in the proceedings of:};
		\node[anchor=west] (B) at (.2,-.5) {\emph{39$^\text{th}$ IEEE/ACM Int. Conf. on Automated Software Engineering}};
		\node[text width=.46\textwidth,anchor=north west] (C) at (0,-.9) {\footnotesize The present document is the preliminary version of the work prior to peer-\\[-1mm] review. The final version can be found on the publisher website.};
		\filldraw[rounded corners=2pt,fill=gray,draw=gray!25,opacity=0.25] (C.south west) rectangle (A.north east); %
	\end{tikzpicture}
\end{center}

\section{Introduction}\label{sec:introduction}

With the exponential growth of web applications and the increasing complexity of software systems, the demand for efficient and reliable testing methodologies has become paramount. Among the various forms of testing, the black-box testing of \emph{REpresentational State Transfer} (REST) APIs has garnered significant attention due to their widespread adoption in modern web architectures.

In literature, several black-box REST API testing approaches have been proposed~\cite{Arcuri-TOSEM23-Survey}, mostly relying on the information available in the \emph{OpenAPI Specification} (OAS), that is the API formal documentation of the system under test. However, it has been observed that this information often falls short of addressing the intricate challenges associated with REST API testing. Specifically, two primary challenges arise in black-box REST API testing: \textit{(i)}~the selection of an effective ordering of API operations to test; and \textit{(ii)}~the generation or retrieval of valid input values for such operations.

In contrast to web or mobile application testing, where the next test interaction can often be deduced from the application's context (\eg available links or widgets in the graphical user interface), REST APIs expose no such explicit context. Indeed, API operations can be, in principle, called at any time, and the OAS usually does not encode any information about the prioritization of operation invocations. %
This may lead to failures in crafting test cases, not just due to incorrect generation of operation input values but due to calling an operation that is not (yet) ready to be called, given the current API state. 

As a simple example, consider the case of an e-commerce service. When the shopping cart is empty, the checkout operation is supposed to fail independently of the input values provided to the operation simply because the cart is not ready for checkout. This suggests that an operation adding items to the cart must be called first to make the cart available for checkout. Some state-of-the-art black-box testing tools~\cite{Atlidakis2019RESTler,Corradini2022,morest} alleviate the problem of operation ordering by considering producer-consumer relations between API operations. This yields an operation ordering purely based on explicit data dependencies derivable from the OAS (that is, an operation is tested before another when the latter requires as input a resource provided as output by the former). Even if effective in simple cases, such an approach overlooks \emph{implicit} operation dependencies caused by \emph{logic constraints} rather than \emph{data constraints}, as in the case of the checkout example above. Conversely, \emph{spurious constraints} could be incorrectly inferred from the OAS,  driving testing tools towards inappropriate ordering of operation sequences. Indeed, an OAS does not fully encode the API business logic, resulting in API constraints that cannot be statically inferred from it. %

In addition to the correct operation ordering, another crucial goal when testing REST APIs is adopting valid input values to call operations. This task is particularly hard due to the extensive value exploration required within a vast input space. Valid input values can encompass various combinations of data, which may need to satisfy inter-parameter dependencies and value conditions. Some input constraints are documented in the OAS and exploited by testing tools~\cite{Atlidakis2019RESTler,Atlidakis2020SecurityProperties,Corradini2022,MartinLopez2021DependenciesAPI,kim2023adaptive} to craft correct HTTP requests. However, similarly to the operations dependency case, some implicit input constraints may not be syntactically retrieved from the OAS, decreasing the chances of crafting successful test cases. For instance, an operation adding products to the cart may fail if a discounted product is added in a quantity less than the minimum. Such constraint is known to the API developer but unlikely to be found in the OAS and, consequently, hidden from black-box testing tools. %
Moreover, smartly picked input values might bring a REST API to a new state that is interesting to test because it could expose new and more complex logic defects.

With the aim of automatically generating effective test cases for REST APIs, we propose \deeprtg, a novel approach based on \emph{deep reinforcement learning} to learn API constraints during testing, potentially even those not documented in the OAS (hence hidden from black-box tools). \deeprtg trains an intelligent agent to autonomously learn and optimize a strategy to test a REST API in a black-box fashion. Deep reinforcement learning is leveraged to guide an agent in the \emph{exploration} of several API states, which is positively rewarded when discovering an effective order in which to test API operations. At the same time, \deeprtg also learns the most effective strategies to generate input values among the available strategies. Typical strategies are random generation, dictionary lookup, and reusing examples from the OAS.
Such selection is based on the \emph{experience} gained during previous successful interactions. Finally, \deeprtg also performs \emph{exploitation}, by mutating successful interactions and improve test coverage, fault detection, and collect even further experience.

The contribution of this paper can be summarized as follows:
\begin{itemize}
	\item The first approach leveraging deep reinforcement learning to automatically learn an effective \emph{testing order} for operations in a REST API; %
	\item A novel reinforcement learning-based approach to select the most effective \emph{input value generation strategy} for operation input parameters; %
	\item Empirical results demonstrating that the proposed approach is \emph{effective} (coverage and faults detection) and \emph{efficient} (number of requests) in testing REST APIs. Indeed, \deeprtg achieves superior performance than state-of-the-art testing tools a set of case study APIs; %
	\item An open-source tool implementing the approach, that can be found in the replication package~\cite{ReplicationPackage}.%
\end{itemize}

\section{Background}\label{sec:background}

This section covers the background notions needed to understand our approach. It includes an introduction to REST APIs and OpenAPI specifications, automated REST API testing guided by data dependencies, and reinforcement learning.

\subsection{REST APIs and OpenAPI Specifications}
The REST (REpresentational State Transfer) architectural style~\cite{Fielding2000ArchitecturalStyle} is nowadays the most common paradigm adopted in web API development. A RESTful API (or REST API) is a web API that adheres to such a paradigm, allowing web clients to access and manipulate resources and invoke remote routines by leveraging stateless operations over the HTTP protocol.  

\newsavebox{\mybox}

\begin{lrbox}{\mybox}
	\begin{minipage}{\linewidth}
		\begin{lstlisting}[language=yaml,numbers=none,xleftmargin=-7pt,xrightmargin=0pt,multicols=2,breaklines=true,frame=tb,postbreak=\mbox{\textcolor{green}{$\hookrightarrow$}\space},columns=fullflexible,framexleftmargin=5pt]
openapi: "3.0.0"
info:
  version: 1.0.0
  title: "Simple eComm"
  license:
    name: MIT
servers:
  - url: http://simple.ecommerce.io/v1
paths:
  /addProductToCart:
    post:
      summary: "Add product(s) to the cart"
      operationId: addProductToCart
      parameters:
        - name: productId
          in: query
          required: true
          schema:
            type: integer
            format: int64
        - name: quantity
          in: query
          schema:
            type: integer
            default: 1
            minimum: 1
            maximum: 100
      responses:
        '200':
          description: "Product(s) added"

  /products/search:
    get:
      summary: "Search products by name"
      operationId: productSearch
      parameters:
        - name: keyword
          in: query
          required: true
          schema:
            type: string
      responses:
        '200':
          schema:
            type: array
            items:
              properties:
                productId:
                  type: integer
                  format: int64
                name:
                  type: string
                price:
                  type: number
                  format: float
  /checkout:
    post:
      summary: "Finalize purchase"
      operationId: checkout
      responses:
        '200':
          description: "Purchase completed"
		\end{lstlisting}
	\end{minipage}
\end{lrbox}

REST APIs provide a uniform interface to \emph{Create}, \emph{Read}, \emph{Update}, and \emph{Delete} (CRUD) resources, where an HTTP URI identifies a resource while CRUD operations are typically mapped to the HTTP methods \inlineyaml{POST}, \inlineyaml{GET}, \inlineyaml{PUT} (or \inlineyaml{PATCH}) and \inlineyaml{DELETE}, respectively. Additionally, REST APIs can expose functionalities such as resource search, invocation of remote routines, and authentication mechanisms. Upon receiving and processing an HTTP request that exercises a specific API operation, the REST API returns an HTTP response with the outcome of the request, called status code (\eg \twoxx for a success; \fourxx for a client-side error; or \fivexx for a server-side error), and, possibly, a payload.

\begin{figure}[t]
	\hspace*{7pt}
	\scalebox{.94}{\usebox{\mybox}}
	\caption{OpenAPI specification excerpt for \emph{Simple eComm}.}\label{fig:openapiexample}
\end{figure}

As an example, consider \emph{Simple eComm}, a REST API managing a simple e-commerce service. A possible HTTP URI pointing to the search products functionality could be \inlineyaml{/products/search}. In this case, the HTTP operation \inlineyaml{GET /products/search} is used to search products by name, listing all products matching the search keyword provided as an HTTP input parameter. Conversely, the HTTP operation \inlineyaml{POST /addProductToCart} could be used to add a product to the shopping cart.

REST APIs are usually documented by using the OpenAPI\footnote{\url{https://www.openapis.org/}} standard. According to such standard, an API is described by a structured file (either YAML or JSON), called \emph{OpenAPI Specification} (OAS), that indicates how to reach the API using a URI, which authentication schema is adopted, and the details of the API available operations: the input parameters (and their schema) to be used in requests and the schema of responses. 

Figure~\ref{fig:openapiexample} contains an excerpt of the OAS for \emph{Simple eComm}. After an initial header that specifies versions, licenses, and the base URL of the REST API, an OpenAPI specification contains the list of available URL paths. In the example, we have, among others, the two paths \inlineyaml{/products/search} and \inlineyaml{/addProductToCart}. %

A REST API \emph{operation} is a pair of path and HTTP method, usually identified by an operation identifier. For instance, the method \inlineyaml{GET} in \inlineyaml{/products/search} refers to the operation identified as \inlineyaml{productSearch}, %
where the search string is given as query parameter \inlineyaml{keyword}. Similarly, the method \inlineyaml{POST} in \inlineyaml{/addProductToCart} refers to the operation named \inlineyaml{addProductToCart}, %
where the product to add and quantity of the added product are given as query parameters \inlineyaml{productId} and \inlineyaml{quantity}, respectively. %

Request input and output are associated with a schema that specifies their type and, optionally, a set of constraints on values (\eg a minimum or a maximum value for numeric parameters as in the case of \inlineyaml{quantity}). Types can be atomic (\eg integers and strings) or structured (\ie compound objects). For instance, the parameter \inlineyaml{keyword} of \inlineyaml{/products/search} is of type \inlineyaml{string}, while the response to the corresponding \inlineyaml{GET} operation is expected to be an array of product objects, whose schema is also provided in the specification. %

\subsection{Data Dependency-based REST API Testing}

Testing strategies of black-box approaches are typically based only on the information contained in OpenAPI specifications. So, to select an effective ordering of operations to test, state-of-the-art tools~\cite{Atlidakis2019RESTler,Corradini2022,morest} purely base their decisions on data dependencies among the documented operations. 

For instance, \rler~\cite{Atlidakis2019RESTler} infers data dependencies (in the form of producer-consumer relations) between the operations documented in the OAS. Then, by leveraging a search-based algorithm, it extensively generates sequences of HTTP requests conforming to the inferred dependencies. Instead, \rtg~\cite{Corradini2022} computes the \emph{Operation Dependency Graph} (ODG), a graph encoding data dependencies (again, producer-consumer relations) among operations available in the OAS. Finally, Morest~\cite{morest} exploits the \emph{Property Graph}, which captures OAS-induced relations between API operations, to prioritize operation testing order.
 
Referring to the \emph{Simple eComm} example of Figure~\ref{fig:openapiexample}, the operation \inlineyaml{addProductToCart} depends on the operation \inlineyaml{productSearch}, which provides a list of valid products, since the output of the latter can be used as input for the former. 
Indeed, to test the operation \inlineyaml{addProductToCart}, a valid \inlineyaml{productId} value is needed that is unlikely to be guessed. Hence, the operation \inlineyaml{productSearch} should be tested earlier in order to fetch a valid value for \inlineyaml{productId}. %

Data dependencies are statically inferred from the OAS by matching parameter names and schemas, giving higher priority to operations with satisfied data dependencies. %

\subsection{Reinforcement Learning} 
Reinforcement learning is a paradigm of machine learning in which an agent learns to make decisions by interacting with an environment. After taking action, an agent receives feedback in the form of rewards or penalties depending on the effect of its actions and adjusts its strategy over time to maximize cumulative rewards. The learning process involves discovering an optimal policy that guides decision-making.

\textbf{Multi-Armed Bandit Problem.} The multi-armed bandit problem is a decision-making scenario where an agent is confronted with a problem, where either exploration or exploitation should be aimed at the same time. This problem is named after its origin, where a set of slot machines is available to a player with a limited playing budget. Each slot machine has an unknown probability distribution and amount of winning, so the player has to spend money to \emph{explore} the machines to find the best to play with but, at the same time, money should be invested to \emph{exploit} best machines to gain profit. 
The challenge lies in the exploration vs. exploitation trade-off: the agent must strike a balance between trying different actions to uncover their reward potentials and exploiting the current best-known action for immediate gains. This dilemma reflects the tension between gaining more information and making optimal decisions based on existing knowledge.

A class of solutions to the multi-armed bandit problem, known as \emph{probability matching strategies}, is based on the idea that the probability of choosing a solution should be equal to the probability for that solution to be optimal, according to the collected experience.

\textbf{Deep Reinforcement Learning.} %
Deep reinforcement learning combines reinforcement learning with deep neural networks, using the latter to represent the agent's policy or value function. The learning process revolves around the agent interacting with an environment, where it observes a state $s_t$, takes an action $a_t$, receives a reward $r_t$, and transitions to a new state $s_{t+1}$. The agent's goal is to learn a policy that maximizes the cumulative reward over time.

The \emph{Proximal Policy Optimization} (PPO) algorithm~\cite{ppo} is a technique used in deep reinforcement learning to train an agent's decision-making abilities. PPO aims to improve an agent's policy through trial and error, balancing the policy update towards better performance while keeping it similar to the previous policy. %
This balance helps the agent learn effectively while maintaining stability.

\section{Motivating Example}\label{sec:motivating-example}
Let us consider the OAS of \emph{Simple eComm} reported in Figure~\ref{fig:openapiexample}. In the excerpt, three operations are defined: \inlineyaml{productSearch}, which searches for products by name; \inlineyaml{addProductToCart}, which adds a product to the shopping cart; and \inlineyaml{checkout}, which finalizes the purchase of the current shopping cart. Even if apparently simple, by using \emph{Simple eComm} we can highlight three non-trivial challenges posed by REST APIs, that are only partially addressed by state-of-the-art testing approaches.

\emph{Challenge 1: Correct ordering of operations to test.}  
To successfully test the \inlineyaml{checkout} operation, we need a non-empty shopping cart, and this requires: \emph{(i)} searching for existing products; \emph{(ii)} adding them to the shopping cart; and \emph{(iii)} performing the checkout. %

State-of-the-art testing approaches based on data dependencies %
infer the ordering on which operations are tested by exploiting producer-consumer relations. %
The (explicit) data dependency between \inlineyaml{productSearch}, that returns a \inlineyaml{productId} as output, and \inlineyaml{addProductToCart}, that requires a \inlineyaml{productId} as input, would suggest testing \inlineyaml{addProductToCart} after \inlineyaml{productSearch}, using the \inlineyaml{productId} returned by the latter. However, such approaches would overlook the business logic constraint that only valid and non-empty carts can be finalized. This results from an \emph{implicit dependency} between \inlineyaml{addProductToCart} and \inlineyaml{checkout}, where the former operation takes the state where it can be purchased, and the latter operation finalizes the purchase. This dependency cannot be inferred just from the OAS of \emph{Simple eComm} in Figure~\ref{fig:openapiexample}, but it could be learned by testing the API.
We advocate for the need to overcome the limitations of current approaches and enhance automated testing so that implicit dependencies are learned on successfully tested scenarios, yielding an effective ordering of operations to test.
	
\emph{Challenge 2: Appropriate input data selection.} 
The second limitation is in how state-of-the-art approaches decide input data. Typical strategies are either random generation according to the constraints in the API specification, sourcing values from it (\eg default, enums, and example values), or reusing values observed on previous interactions (from HTTP responses). However, the strategy to be used is typically chosen randomly among those available, regardless of the role of the input parameter. For instance, the value of \inlineyaml{productId} parameter in the \inlineyaml{addProductToCart} operation may be wisely chosen by picking one of those values that have been observed in previous responses (\eg to \inlineyaml{productSearch}), rather than with random generation. Conversely, it is more effective to use a random query string to test \inlineyaml{productSearch}, rather than previously observed item values. We advocate that the decision about the strategy to generate input data should exploit the \emph{experience} on past successful interactions.

\emph{Challenge 3: Balance between exploration and exploitation.} 
Lastly, a notable challenge is how to balance \emph{exploration} of the REST API to test new operations and \emph{exploitation} of already tested operations for increasing  coverage by testing them with diverse inputs. 
In fact, %
thorough and in-depth testing of individual operations could yield two distinct benefits: \emph{(i)} the acquisition of additional valid data from API responses and \emph{(ii)} higher coverage of API source code.

In the context of the \emph{Simple eComm} example, intensifying the testing on \inlineyaml{searchProduct}, by trying different search keywords, would increase the chances of formulating valid queries and retrieving more and more shopping products. Additionally, having multiple valid \inlineyaml{productId} values would facilitate the testing of the subsequent operation \inlineyaml{addProductToCart} and, consequently, of the operation \inlineyaml{checkout}.

\begin{figure}[t]
	\centering
	\input{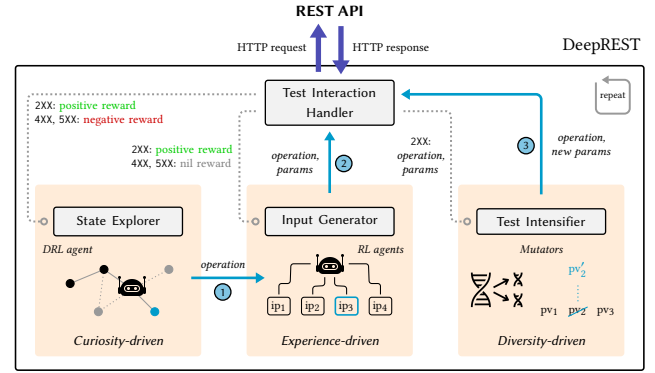}
	\caption{Approach overview.}\label{fig:approach-overview}
\end{figure}

\section{Approach}
\label{sec:approacj}
In this section we present \deeprtg, our approach conceived to address the challenges identified in the previous section. 
An overview of the architecture of \deeprtg is illustrated in Figure~\ref{fig:approach-overview}, depicting an iterative process that applies three main components to craft REST API test cases, ending up in actual HTTP interactions. Such an iterative process is repeated until all API operations have been extensively tested or the allocated testing budget expires.

The first component, the \emph{Curiosity-Driven State Explorer}, is a deep reinforcement learning agent determining the most relevant next API operation to test based on the current state of the API. Employing a curiosity-driven strategy, this agent is positively rewarded when it reaches new API states, thereby \emph{promoting exploration}. Its objective is to cover new states of the API beyond those reachable purely following data dependencies, discovering those that are reachable when implicit dependencies are exploited. Further details about this component are described in Subsection~\ref{sec:curiosity-driven-state-exploration}.

The second component, the \emph{Experience-Driven Input Generator}, is responsible for supplying valid input data to operations. Internally, this component adopts a group of reinforcement learning agents, each dedicated to a distinct input parameter. Treating the parameter value selection as a multi-armed bandit problem, these agents \emph{leverage experience} to make informed decisions. Experience is also exploited to decide whether to include non-mandatory parameters in requests and to determine arrays' size. Further details about this component are provided in Subsection~\ref{sec:experience-driven-input-generation}.

The first two components collectively assemble an HTTP request for the API under test. Testing an API is learned by providing a positive reinforcement to correct interactions (HTTP responses with a status code \twoxx) and a negative reinforcement to wrong interactions (HTTP responses with a status code \fourxx or \fivexx).

Upon successfully reaching a new API state and testing a new operation, a \emph{test intensification phase} is initiated. The third component, the \emph{Mutation-Based Test Intensifier}, performs deeper testing of the newly reached state, with the twofold objective of collecting additional test data from API responses and increasing test coverage. Intensification consists of applying mutation operators to a successful HTTP request to generate many new requests, each with small changes in parameter values, \emph{promoting input diversity}. Further details about this component are described in Subsection~\ref{sec:mutation-based-exploitation}.

\subsection{Curiosity-driven State Exploration}\label{sec:curiosity-driven-state-exploration}
To comprehensively inspect the behavior of the API under test, it is crucial to invoke API operations in a clever order. As pointed out in Section~\ref{sec:motivating-example}, if the order in which operations are executed is based on data dependencies only, we might overlook potential \emph{implicit dependencies} and, thus, fail to test some operations. Moreover, {\em spurious dependencies} could be inferred on the OAS, causing a testing approach to waste its testing budget on hopeless attempts.
This limitation could result in a partial exploration of the API's state space, collecting few meaningful test data and causing low test coverage. On the flip side, it is not advisable to test all possible sequences of operations since %
the sheer volume of potential sequences is prohibitively vast. In addition, a substantial portion of these sequences lacks practical significance or coherence with the API business logic, making their testing unnecessary. 
Hence, a selective approach is crucial for efficient and effective testing.

Leveraging curiosity-driven deep reinforcement learning, we push toward exploring the API while learning an effective order of operations. 
We formulated the problem as an instance of reinforcement learning, with proper {\em state}, {\em actions}, {\em state transition}, and {\em reward} as follows.

\textbf{State.} In an ideal scenario, the \emph{DRL state} should precisely reflect the internal state of the API under test. However, since we treat the API as a black box, its precise internal state is unknown. We then resort to an approximation of the API internal state, referred to as \emph{state observation}, derived from the information we can retrieve from previous HTTP interactions with the API. For instance, a successful exercise of the \inlineyaml{addProductToCart} operation inherently suggests that the shopping cart contains items. With this consideration, we represent the DRL state as a list of $n$ integers, where $n$ is the number of API operations. The value of the element at position $i$ in the list represents the number of successes (HTTP interactions that returned a \twoxx status code) for the $i$-th operation in the OAS that occurred so far during testing. %

As an example, consider the three operations of \emph{Simple eComm}: \inlineyaml{addProductToCart}, \inlineyaml{productSearch}, and \inlineyaml{checkout}. Then, the list $[\,0~1~0\,]$ represents a state observation where the operation \inlineyaml{productSearch} has been successfully tested (1 success), while the other two operations might have been attempted with no success (0 successes). %
In fact, successful operations are supposed to alter the (internal) state of an API, potentially enabling other operations to be called, while failed operations should have no side-effects.

Considering that a state observation is multi-discrete, \ie it is an array with discrete values, finite-state reinforcement learning techniques, such as Q-Learning~\cite{Watkins1992} can not be applied, and \emph{deep} reinforcement learning is needed. %

\textbf{Actions.} The agent actions are represented by the set of available API operations. In our example, the three actions are testing either \inlineyaml{addProductToCart}, \inlineyaml{productSearch}, or \inlineyaml{checkout}.

\textbf{State Transitions.} The API transitions to a different (internal) state depending on the outcome of an operation's execution. If an operation execution resulted in success (HTTP status code \twoxx) the API is assumed to have reached a new state, potentially ``enabling'' operations that were infeasible to test in the previous state. In this case, we update the state observation by incrementing the success counter for the operation successfully tested. Conversely, in the event of failed interactions (HTTP status codes \fourxx and \fivexx), we assume that the API state did not change. In this case, also the state observation does not change.

\textbf{Reward.} To stimulate curiosity, a DRL agent usually receives a large positive reward whenever its action takes the environment to a new state never reached so far~\cite{Zheng-ICSE-2021,Pan-ISSTA-2020}. In our context, new (internal) API states are approximated with new state observations, which are reached when the DRL agent is able to successfully test an operation. Here, we distinguish two cases: the operation is successfully tested for the first time (\ie in the previous state observation its counter was $0$); and the operation has been already successfully tested before (\ie in the previous state observation its counter was greater than $0$). In the first scenario, the agent receives a large positive reward ($+1000$), while in the second scenario, the agent receives a negative reward ($-100$). With this strategy, we stimulate curiosity and, at the same time, we do not overly discourage the agent from repeating certain operations, which may be a prerequisite for successfully testing other operations.

Finally, if the API rejects the chosen operation (with a \fourxx or \fivexx status code), a slightly negative reward ($-1$) is assigned to the agent. The penalty is intentionally mild, recognizing that rejection can stem from various reasons. Indeed, the agent may have selected the correct operation, but value generation provided the wrong input data (the latter causing the failure). In such a case, the agent should not be discouraged from attempting again to test the operation with different input data. Recall that input generation is in charge of a different learning agent (that will be presented in Section~\ref{sec:experience-driven-input-generation}), having a different reward strategy.

\textbf{DRL Algorithm.} To solve the aforementioned learning problem, we opted to employ the \emph{Proximal Policy Optimization} algorithm~(PPO)~\cite{ppo} since it is one of the most recent and advanced deep reinforcement learning algorithms~\cite{openai} supporting vector state space with discrete values. In particular, we employed the PPO implementation coming with Stable Baselines 3~\cite{raffin2021stable}. We customized the episode length to be equal to $\smalltt{ep\_length} = 20 \cdot \smalltt{num\_operations}$, thus correlating it with the size API being tested. APIs exposing several operations will be granted more attempts in a single episode than smaller APIs. In a similar fashion, the maximum value for counters in the DRL state is set to $20$, and in the case that an operation is successfully tested more than 20 times in an episode, the episode is truncated before its natural end, as the DRL agent could have stacked in choosing simple operations to test, rather than exploring others. %

\subsection{Experience-driven Input Generation}\label{sec:experience-driven-input-generation}
Providing appropriate input data in API requests is crucial in automated black-box testing of REST APIs. Such input data comprises a collection of operation parameters that can take the form of simple values (such as strings, numbers, and booleans), arrays, or compound objects. In the process of generating input data, three challenges must be addressed: \emph{(i)} deciding which parameters to include in the request among the optional ones; \emph{(ii)} selecting the length of arrays; and, most notably, \emph{(iii)} deciding the values to assign to primitive-type parameters (recursively on compound objects).

To address these challenges, we developed a novel approach based on cumulative experience inspired by the multi-armed bandit problem. A distinct agent is deployed for each parameter, responsible for making clever decisions about how that parameter should be used in requests, grounded in their accumulated experience. 
Agents' experience is initially empty, thus leading to initial purely random decisions. When successful input data is generated, agents receive a positive reward ($+1$) associated with the decisions they made which led to a successful interaction, accumulating valuable experience on how to succeed. Decisions are made following a probability-matching strategy, where the likelihood of making a specific decision corresponds to the observed statistical distribution estimated at that time, as follows. 
The probability of choosing $d$ is equal to the ratio of the accumulated rewards for $d$ (\ie $R_d$) on the total rewards $\sum_i{R_i}$ accumulated so far, that is: $\mathbb{P}(d) = \nicefrac{R_d}{\sum_i{R_i}}$.

For example, consider an agent deciding whether to include a parameter in a new request. Suppose that, based on the agent's accumulated experience, the parameter has been employed in $8$ out of the $10$ previously successful interactions (\ie it has collected $8$ reward points). The agent will incorporate that parameter in the new request with a probability of $\mathbb{P} = 0.8$. To prevent over-fitting towards a single solution, however, agents might also make random decisions (with low probability, \eg $0.1$), still allowing the exploration of not yet tried configurations.%

A notable aspect of our approach is that the experience garnered on a parameter in one operation is leveraged for making decisions on parameters with the same (or similar) name in other operations. This approach significantly reduces the learning time when testing new operations, promoting efficiency and knowledge reuse.

We now describe how we specifically address the three challenges mentioned at the beginning of the subsection.

\textbf{Parameter Presence.} For each optional parameter in an operation, the agent decides whether to include or exclude the parameter in the request. As already discussed, the probability of a parameter being included matches the statistical distribution estimated so far.

\textbf{Array Length.} Deciding the length of an array is inherently a hard problem, as arrays can theoretically have any size. We categorize array length into three classes to overcome the issue, yielding a bounded approach. In particular, \emph{Class~A} denotes empty arrays (the length is $0$), \emph{Class~B} denotes one-element arrays, and \emph{Class~C} denotes arrays having at least 2 elements. Agents are rewarded positively ($+1$) if their selected size class leads to a successful request. When crafting new requests, agents will select the most suitable size class from the statistical distribution estimated so far. In case the \emph{Class C} is selected, the actual size of the array is randomly chosen to be equal to or greater than 2 and compatible with the length constraints reported in the OAS.

\textbf{Input Values.}
Our approach relies on a catalog of input generation strategies, or \emph{sources of values}, that are wisely selected at testing time to retrieve the most appropriate value for parameters at a given point in testing time. %

Input values could be taken from various sources, such as random generators, examples in the OAS, and dictionaries containing test data collected from previous HTTP interactions. However, the same source might not be equally effective for all the parameters. %
For instance, when dealing with resource identifiers, dictionaries are likely to be the optimal sources of valid values observed in the past rather than random values that are unlikely to be valid identifiers. Our approach accumulates and subsequently leverages knowledge about the most successful sources for each parameter. 

When testing an operation, we keep track of the source that supplied the value for each parameter. Upon successful execution of an operation (status code \twoxx),  reinforcement learning agents obtain a positive reward ($+1$), promoting the reuse of the same source in the future. It is important to note that the same source is unlikely to provide the same, identical value for future requests (think of, for instance, the random generation source). However, sources employ the same strategy to select a value, which ensures further exploration of the API.

The value sources now available in \deeprtg are the following.

\inTextCodeBf{Random}~ Parameter value is randomly generated according to the constraints in the OAS.

\inTextCodeBf{Default}~ Parameter value is the default value in the OAS.

\inTextCodeBf{Enum}~ Parameter value is randomly taken from one of the valid enum values from the OAS.

\inTextCodeBf{Examples}~ Parameter value is assigned with one of the example values from the OAS.

\inTextCodeBf{ResponseDictionary}~ Parameter value is taken from a dictionary of API response values observed for this parameter in previous interactions. These values, coming directly from the API, are likely valid and, therefore, likely accepted in new requests.

\inTextCodeBf{LastResponseDictionary}~ Same as the previous, but the last observed value is assigned to the parameter. This increases the likelihood of the value being valid. In the case of parameters acting as resource identifiers, it will help the generation of a chain of operations targeting the same resource.

\inTextCodeBf{RequestDictionary}~ Parameter value is taken from a dictionary of values already used for this field, whose API requests obtained a successful status code. These values are likely valid since the API has accepted them in previous interactions.

\inTextCodeBf{LastRequestDictionary}~ Same as the previous, but the last observed value is assigned to the parameter.

\inTextCodeBf{LargeLanguageModelDictionary} A large language model is queried to supply values for parameters based on their names, the \textit{context} of the endpoint, and the parameter description if provided in the OAS (further details in the next subsection).

Except for \smallsf{LargeLanguageModelDictionary}, which is a contribution of the paper, the previous value sources are inspired by literature~\cite{Corradini2022,Atlidakis2019RESTler}, and this catalog can be easily extended. The novel contribution of our approach is how to learn the most effective strategy from the catalog at each testing step rather than the catalog itself.

\textbf{Large Language Model Dictionary.} A new method for generating realistic parameter values in our implementation of \deeprtg utilizes the inherent knowledge of large language models.
We assume that a large language model can suggest realistic values for a parameter based on both its name (for instance, for a parameter named \inlineyaml{title}, suggestions could include "The Odyssey" or "The Iliad") and the context of the specific API operation to which the parameter belongs (for example, within the operation \inlineyaml{POST /person}, plausible values for a parameter named \inlineyaml{title} could be, instead, "Mrs." or "Mr."). Furthermore, the natural language descriptions associated with the parameter and its API operation, sometimes found in the OAS, can further guide the language model in generating realistic values.

To implement this approach, before actually starting with the testing session, we ask an LLM to suggest plausible input values for all parameters in the OAS. This involves providing the LLM with the HTTP method and path of the API operation, the description of the API operation (if available in the OAS), the parameter name, its type, and its description from the OAS (if available). The language model then responds with a set of realistic values which are stored in a dictionary for subsequent use while generating tests. 
Practically, we deployed a local instance of GPT4All~\cite{GptForAll} with the model \smalltt{wizardlm-13b-v1.2.Q4\_0.gguf}, described on the GPT4All website as the ``best overall larger model''. For each parameter in the OAS of an  API, the LLM is asked to provide at least 20 relevant values.

\subsection{Mutation-based Test Intensification}
\label{sec:mutation-based-exploitation}

During test case generation, which is guided by curiosity and experience, the moment a new operation is successfully tested for the first time, we shift from exploration to exploitation (or intensification).

This phase involves multiple replays of the successful HTTP request after it has been modified by applying a catalog of mutation operators. Some of them are \emph{nominal mutators}, which alter the initial request while still satisfying all the constraints in the OAS (thus generating further potentially valid requests), while others are \emph{error mutators}, which alter the request in a way that violates some constraints from the OAS (thus generating potentially invalid requests). We believe that an invalid request, resembling a valid one, holds the potential to traverse unexplored branches in the API source code, thereby contributing to increased test coverage and hopefully leading to more effective test cases. 
To build the catalog, we started with mutation operators from literature: some of them keeping HTTP requests valid~\cite{kim2023enhancing}; while others turning HTTP requests invalid~\cite{Corradini2022} (according to the OAS). We then added additional mutation operators such that each parameter of each operation can be changed once to keep the overall request valid and once to turn it invalid. 
Currently, 10 mutation operators are available in \deeprtg (4 nominal mutators and 6 error mutators), described in the following.

\inTextCodeBf{AddParameter}~ An optional parameter is added to the original request. [\emph{nominal}]

\inTextCodeBf{RemoveParameter}~ An optional parameter is removed from the original request. [\emph{nominal}]

\inTextCodeBf{RefillValue}~ The value of a parameter is updated with a new value that satisfies the constraints in the OAS. [\emph{nominal}]

\inTextCodeBf{NumberBoundaries}~ In the case of a numeric parameter, its value is changed to be close to its boundaries. [\emph{nominal}]

\inTextCodeBf{AddInvalidParameter}~ An optional parameter with an invalid value is added to the original request. [\emph{error}]

\inTextCodeBf{NumberOutOfBoundaries}~ In the case of a numeric parameter, its value is changed to out of its boundaries. [\emph{error}]

\inTextCodeBf{ChangeHttpMethod}~ The HTTP method of the original request is changed. [\emph{error}]

\inTextCodeBf{MissingRequired}~ A mandatory parameter is removed from the original request. [\emph{error}]

\inTextCodeBf{WrongType}~ A parameter value is replaced with a new one of a different type. [\emph{error}]

\inTextCodeBf{ConstraintsViolation}~ A parameter value is changed so that it violates the constraints from the OAS. [\emph{error}]

\section{Evaluation}\label{sec:evaluation}

In this section, we conduct a thorough empirical evaluation of \deeprtg. Our primary focus is assessing the \emph{test coverage} achieved by our approach and its \emph{fault detection} capability. Test coverage, intended as both code coverage and operations successfully tested, is directly correlated with Challenges 1 and 2 presented in Section~\ref{sec:motivating-example}. Indeed, an optimal testing order of operations (Challenge 1) is likely to result in testing more operations successfully. Similarly, an appropriate and diverse input value selection (Challenge 2) is likely to result in testing more parts of the API implementation, increasing code coverage. Alongside successfully testing operations, to individuate faults a balance between exploration and exploitation (Challenge 3 of Section~\ref{sec:motivating-example}) is needed, testing again operations with different input values potentially inducing server-side errors. Since HTTP interactions are time-consuming, \emph{testing efficiency}, in terms of the number of requests, is also investigated. To validate the aforementioned aspects, we engage in a comparative analysis with the current state-of-the-art testing tools for REST APIs.

\subsection{Research Questions}
To guide our empirical evaluation, we formulated the following two research questions.

\begin{description}
	\item[\textbf{\req{1}}:] What is the \emph{effectiveness} of \deeprtg in generating test cases for REST APIs? How does it compare with state-of-the-art approaches?
	\item[\textbf{\req{2}}:] What is the \emph{effeciency} of \deeprtg in generating test cases for REST APIs? How does it compare with state-of-the-art approaches?
\end{description}

To answer \req{1}, we execute \deeprtg and five state-of-the-art REST API testing tools on a benchmark set of API case studies. For each tool, targeting each API in the benchmark, we measure code coverage, count successfully tested operations (\twoxx), and server-side failures (\fivexx).

To answer \req{2}, for each experiment of the previous research question, we monitor code coverage and success/failure count over time %
to measure their progressive increase. The efficiency of a tool is deemed higher if it achieves high code coverage or success/failure count earlier in the testing process.

\begin{table}[t]
\caption{Benchmark APIs.}
	\label{tab:api-benchmark}
    \resizebox{\linewidth}{!}{%
	\begin{tabular}{l|c|c|c}
		\textbf{API} & \textbf{Short Name} & \textbf{\# Operations} & \textbf{Source} \\ \hline\hline
            REST Countries & rest-countr & 22 & Kim et al.~\cite{kim2023adaptive} \\ %
            User Management & user-mgmt & 22 & Kim et al.~\cite{kim2023adaptive} \\ %
            Market & market & 13 & Kim et al.~\cite{kim2023adaptive} \\ %
            Project Tracking System & proj-track & 67 & Kim et al.~\cite{kim2023adaptive} \\ %
            Features Service & feat-serv & 18 & Kim et al.~\cite{kim2023adaptive} \\ %
            NCS & ncs & 6 & Kim et al.~\cite{kim2023adaptive} \\ %
            SCS & scs & 11 & Kim et al.~\cite{kim2023adaptive} \\ %
            Genome-Nexus & genome-nex & 23 & Kim et al.~\cite{kim2023adaptive} \\ %
            Person Controller & person-ctrl & 12 & Kim et al.~\cite{kim2023adaptive} \\ %
            Blog & blog & 52 & GitHub~\cite{blogApi} \\ %
            LanguageTool & lang-tool &2 & Kim et al.~\cite{kim2023adaptive} \\ %
	\end{tabular}
    }
\end{table}

\subsection{Experiment Setup}\label{ssec:exp-setup}

\textbf{Baseline Testing Tools.} 
Our evaluation consists of a comparison of \deeprtg with state-of-the-art REST API testing tools. We selected a total of five other tools. They are \rtg~\cite{Corradini2022}, Morest~\cite{morest} and Restler~\cite{Atlidakis2019RESTler} as representative of data dependencies-based approaches, \aratrl~\cite{kim2023adaptive} as the only existing tool based on reinforcement learning (although using the non-deep algorithm Q-Learning), and Schemathesis~\cite{schemathesis} as a recently proposed tool whose performance seems competitive~\cite{kim2023adaptive}. According to recent surveys~\cite{kim2023enhancing,kim2023adaptive}, \aratrl and \rtg appear to be the best-performing black-box tools.

\textbf{API Case Studies.}  
To establish a benchmark for our evaluation, we selected a total of 11 API case studies, which are reported in Table~\ref{tab:api-benchmark}. We included all the 10 APIs sourced from a recent study~\cite{kim2023adaptive}. To enhance the realism of our evaluation, we supplemented this set with
an additional API from GitHub (i.e., {\em Blog}) as representative of more complex APIs. In fact, it contains implicit operation dependencies that cannot be syntactically deduced from the OAS (we found such implicit dependencies by manually reverse engineering the API business logic). %
Table~\ref{tab:api-benchmark} reports the list of API case studies, accompanied by the number of operations defined in their respective OAS.

\textbf{Metrics.}  
We gathered the following metrics throughout all testing sessions conducted with all tools and configurations.

\smallsf{Code Coverage}~ %
We collected method, branch, and line coverage to measure the extent of API code executed by the test cases.%

\smallsf{Operation Coverage}~ This black-box coverage metric~\cite{MartinLopez2019TestCoverage,corradini2021restats} is meant to count the number of successfully tested API operations by tools with respect to the total number of operations defined in the API documentation.  %

\smallsf{Fault Detection}~ As a measure of fault detection capacity, we counted the number of server-side failures (status code \fivexx) with unique error messages identified by tools. An error message is considered unique when it is sufficiently different from other messages, as previously defined by Kim et al.~\cite{kim2023adaptive}.

\smallsf{Area Under Curve}~ As an efficiency metric, we captured the progress of code coverage, operation coverage, and fault detection while testing. This is measured as the area under the curve in a graph depicting test coverage and fault detection trends.

\textbf{Experimental Procedure.} 
Experiments have been run using Docker containers, with a separate container for each testing tool and for each case study. Each container is assigned a maximum of 8 cores and 16 GB of RAM. At the end of each experiment run, containers were stopped and rebooted with a fresh file system to avoid any side effects, either in tools or API case studies.
Testing tools were executed with the same budget of API calls for each API in the benchmark. This ensures that all the tools have the same opportunity to explore and test the APIs, regardless of how long an API can non-deterministically take to respond. The budget has been computed for each API independently by empirically checking the amount of interactions that each API can serve in approximately one hour of testing.
To control the impact of non-deterministic features of testing tools, all experiments have been repeated 10 times, reporting the average results. JaCoCo~\cite{jacoco} was deployed to collect source code coverage, while Restats~\cite{corradini2021restats} was utilized for computing the operation coverage from HTTP logs. Metrics were recorded every 5 seconds.%

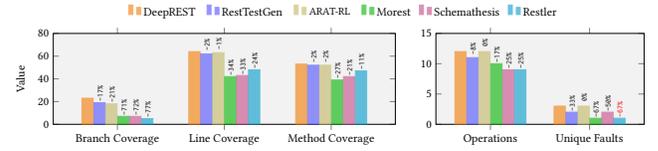
\begin{figure}[t]
\begin{tikzpicture}[scale=.5]
	\begin{axis}[
		ylabel style={yshift=-0.4cm},
		ybar=.5pt,
		axis background/.style={fill=gray!10},
		legend style={at={(.85,1.35)},anchor=north,legend columns=-1,draw=none},
		bar width=.3cm,
		width=.6\textwidth,
		height=4cm,
		symbolic x coords={Branch Coverage,Line Coverage,Method Coverage},
		xtick=data,
		nodes near coords,
		nodes near coords align={vertical},
		ymin=0,ymax=80,
		ylabel={Value\vphantom{gplt}},
		every node near coord/.append style={font=\scriptsize\ttfamily,opacity=1,rotate=90,xshift=7pt,yshift=-5.75pt},
		point meta=explicit symbolic,
		x tick label style={align=center,rotate=0},
		enlarge x limits=.3,
		]
		\addplot[draw=orange!95!black,fill=orange!95!black,opacity=.6] table[x=Metric,y=DeepRest.mean.round,col sep=comma]{data/aggregate/effectiveness-1.csv};
		\addplot[draw=blue!70,fill=blue!70,opacity=.6] table[x=Metric,y=RestTestGen.mean.round,meta=RestTestGen.delta.round,col sep=comma]{data/aggregate/effectiveness-1.csv};
		\addplot[draw=olive!60,fill=olive!60,opacity=.6] table[x=Metric,y=Arat.mean.round,meta=Arat.delta.round,col sep=comma]{data/aggregate/effectiveness-1.csv};
		\addplot[draw=green!70!black,fill=green!70!black,opacity=.6] table[x=Metric,y=Morest.mean.round,meta=Morest.delta.round,col sep=comma]{data/aggregate/effectiveness-1.csv};	
		\addplot[draw=magenta!80!black,fill=magenta!80!black,opacity=.6] table[x=Metric,y=Schemathesis.mean.round,meta=Schemathesis.delta.round,col sep=comma]{data/aggregate/effectiveness-1.csv};
		\addplot[draw=cyan!80!black,fill=cyan!80!black,opacity=.6] table[x=Metric,y=Restler.mean.round,meta=Restler.delta.round,col sep=comma]{data/aggregate/effectiveness-1.csv};
		\legend{\deeprtg\hphantom{S},\rtg\hphantom{S},\aratrl\hphantom{S},Morest\hphantom{S},Schemathesis\hphantom{S},Restler}
	\end{axis}
\end{tikzpicture}
\hspace*{-60pt}
\begin{tikzpicture}[scale=.5]
	\begin{axis}[
		ylabel style={yshift=-0.4cm},
		ybar=.5pt,
		axis background/.style={fill=gray!10},
		bar width=.3cm,
		width=.4\textwidth,
		height=4cm,
		symbolic x coords={Operations,Unique Faults},
		xtick=data,
		nodes near coords,
		nodes near coords align={vertical},
		ymin=0,ymax=15,
		every node near coord/.append style={font=\scriptsize\ttfamily,opacity=1,rotate=90,xshift=7pt,yshift=-5.75pt},
		point meta=explicit symbolic,
		x tick label style={align=center,rotate=0},
		enlarge x limits=.55,
		]
		\addplot[draw=orange!95!black,fill=orange!95!black,opacity=.6] table[x=Metric,y=DeepRest.mean.round,col sep=comma]{data/aggregate/effectiveness-2.csv};
		\addplot[draw=blue!70,fill=blue!70,opacity=.6] table[x=Metric,y=RestTestGen.mean.round,meta=RestTestGen.delta.round,col sep=comma]{data/aggregate/effectiveness-2.csv};
		\addplot[draw=olive!60,fill=olive!60,opacity=.6] table[x=Metric,y=Arat.mean.round,meta=Arat.delta.round,col sep=comma]{data/aggregate/effectiveness-2.csv};
		\addplot[draw=green!70!black,fill=green!70!black,opacity=.6] table[x=Metric,y=Morest.mean.round,meta=Morest.delta.round,col sep=comma]{data/aggregate/effectiveness-2.csv};	
		\addplot[draw=magenta!80!black,fill=magenta!80!black,opacity=.6] table[x=Metric,y=Schemathesis.mean.round,meta=Schemathesis.delta.round,col sep=comma]{data/aggregate/effectiveness-2.csv};
		\addplot[draw=cyan!80!black,fill=cyan!80!black,opacity=.6] table[x=Metric,y=Restler.mean.round,meta=Restler.delta.round,col sep=comma]{data/aggregate/effectiveness-2.csv};
	\end{axis}
\end{tikzpicture}
\caption{Effectiveness results (aggregate).}\label{fig:effectiveness-aggregate}
\end{figure}

\subsection{Experiment Results}

\begin{figure*}[t]

	\begin{tikzpicture}[scale=.5]
		\begin{axis}[
			ylabel style={yshift=-0.5cm},
			ybar=.5pt,
			axis background/.style={fill=gray!10},
			bar width=.2cm,
			width=1.5\textwidth,
			height=4cm,
			legend style={at={(.5,1.35)},anchor=north,legend columns=-1,draw=none},
			symbolic x coords={rest-countr,user-mgmt,market,proj-track,feat-serv,ncs,scs,genome-nex,person-ctrl,blog,lang-tool},
			xtick=data,
			nodes near coords,
			nodes near coords align={vertical},
			ymin=0,ymax=110,
			ylabel={Branch Coverage\vphantom{gplt}},
			every node near coord/.append style={font=\scriptsize\ttfamily,opacity=1,rotate=90,xshift=7pt,yshift=-5.75pt},
			point meta=explicit symbolic,
			xticklabels=\empty,
			enlarge x limits=.05,
			]
			\addplot[draw=orange!95!black,fill=orange!95!black,opacity=.6] table[x=Api,y=DeepRest.mean.round,col sep=comma]{data/apis/branch-coverage.csv};
			\addplot[draw=blue!70,fill=blue!70,opacity=.6] table[x=Api,y=RestTestGen.mean.round,meta=RestTestGen.delta.round,col sep=comma]{data/apis/branch-coverage.csv};
			\addplot[draw=olive!60,fill=olive!60,opacity=.6] table[x=Api,y=Arat.mean.round,meta=Arat.delta.round,col sep=comma]{data/apis/branch-coverage.csv};
			\addplot[draw=green!70!black,fill=green!70!black,opacity=.6] table[x=Api,y=Morest.mean.round,meta=Morest.delta.round,col sep=comma]{data/apis/branch-coverage.csv};
			\addplot[draw=magenta!80!black,fill=magenta!80!black,opacity=.6] table[x=Api,y=Schemathesis.mean.round,meta=Schemathesis.delta.round,col sep=comma]{data/apis/branch-coverage.csv};
			\addplot[draw=cyan!80!black,fill=cyan!80!black,opacity=.6] table[x=Api,y=Restler.mean.round,meta=Restler.delta.round,col sep=comma]{data/apis/branch-coverage.csv};
			\legend{\deeprtg\hphantom{S},\rtg\hphantom{S},\aratrl\hphantom{S},Morest\hphantom{S},Schemathesis\hphantom{S},Restler}
		\end{axis}
	\end{tikzpicture}

	\vspace*{-5pt}
	
	\begin{tikzpicture}[scale=.5]
		\begin{axis}[
			ylabel style={yshift=-0.5cm},
			ybar=.5pt,
			axis background/.style={fill=gray!10},
			bar width=.2cm,
			width=1.5\textwidth,
			height=4cm,
			legend style={at={(.25,.95)},anchor=north,legend columns=-1,draw=none},
			symbolic x coords={rest-countr,user-mgmt,market,proj-track,feat-serv,ncs,scs,genome-nex,person-ctrl,blog,lang-tool},
			xtick=data,
			nodes near coords,
			nodes near coords align={vertical},
			ymin=0,ymax=50,
			ylabel={Operations\vphantom{gplt}},
			every node near coord/.append style={font=\scriptsize\ttfamily,opacity=1,rotate=90,xshift=7pt,yshift=-5.75pt},
			point meta=explicit symbolic,
                x tick label style={rotate=35},
			enlarge x limits=.05,
			]
			\addplot[draw=orange!95!black,fill=orange!95!black,opacity=.6] table[x=Api,y=DeepRest.mean.round,col sep=comma]{data/apis/operations.csv};
			\addplot[draw=blue!70,fill=blue!70,opacity=.6] table[x=Api,y=RestTestGen.mean.round,meta=RestTestGen.delta.round,col sep=comma]{data/apis/operations.csv};
			\addplot[draw=olive!60,fill=olive!60,opacity=.6] table[x=Api,y=Arat.mean.round,meta=Arat.delta.round,col sep=comma]{data/apis/operations.csv};
			\addplot[draw=green!70!black,fill=green!70!black,opacity=.6] table[x=Api,y=Morest.mean.round,meta=Morest.delta.round,col sep=comma]{data/apis/operations.csv};
			\addplot[draw=magenta!80!black,fill=magenta!80!black,opacity=.6] table[x=Api,y=Schemathesis.mean.round,meta=Schemathesis.delta.round,col sep=comma]{data/apis/operations.csv};
			\addplot[draw=cyan!80!black,fill=cyan!80!black,opacity=.6] table[x=Api,y=Restler.mean.round,meta=Restler.delta.round,col sep=comma]{data/apis/operations.csv};
		\end{axis}
	\end{tikzpicture}

	\caption{Effectiveness results (per API).} \label{fig:effectiveness-apis}
\end{figure*}
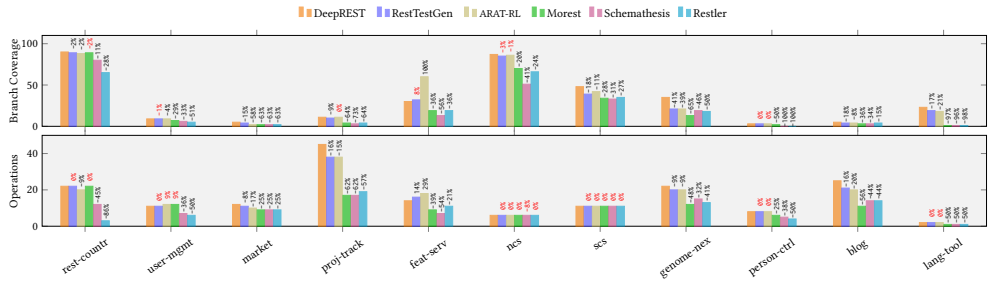

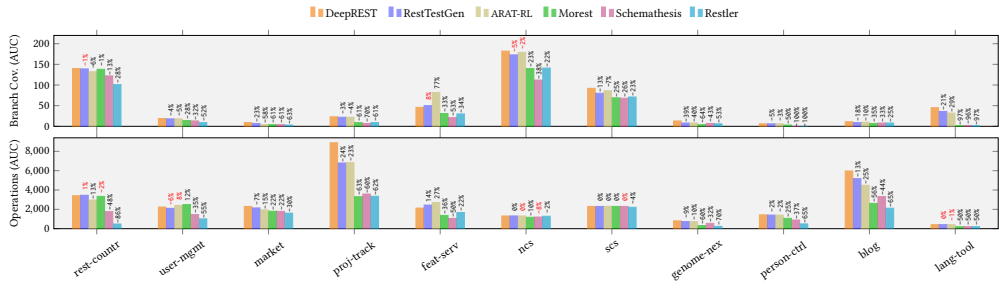
\begin{figure*}[t]

    \begin{tikzpicture}[scale=.5]
    	\begin{axis}[
    		ylabel style={yshift=-0.2cm},
    		ybar=.5pt,
    		axis background/.style={fill=gray!10},
    		bar width=.2cm,
    		width=1.5\textwidth,
    		height=4cm,
    		legend style={at={(.5,1.35)},anchor=north,legend columns=-1,draw=none},
    		symbolic x coords={rest-countr,user-mgmt,market,proj-track,feat-serv,ncs,scs,genome-nex,person-ctrl,blog,lang-tool},
    		xtick=data,
    		nodes near coords,
    		nodes near coords align={vertical},
    		ymin=0,ymax=220,
    		ylabel={Branch Cov. (AUC)\vphantom{gplt}},
    		every node near coord/.append style={font=\scriptsize\ttfamily,opacity=1,rotate=90,xshift=7pt,yshift=-5.75pt},
    		point meta=explicit symbolic,
    		xticklabels=\empty,
    		enlarge x limits=.05,
    		]
    		\addplot[draw=orange!95!black,fill=orange!95!black,opacity=.6] table[x=Api,y=DeepRest.mean.round,col sep=comma]{data/apis/branch-coverage-area.csv};
    		\addplot[draw=blue!70,fill=blue!70,opacity=.6] table[x=Api,y=RestTestGen.mean.round,meta=RestTestGen.delta.round,col sep=comma]{data/apis/branch-coverage-area.csv};
    		\addplot[draw=olive!60,fill=olive!60,opacity=.6] table[x=Api,y=Arat.mean.round,meta=Arat.delta.round,col sep=comma]{data/apis/branch-coverage-area.csv};
    		\addplot[draw=green!70!black,fill=green!70!black,opacity=.6] table[x=Api,y=Morest.mean.round,meta=Morest.delta.round,col sep=comma]{data/apis/branch-coverage-area.csv};
    		\addplot[draw=magenta!80!black,fill=magenta!80!black,opacity=.6] table[x=Api,y=Schemathesis.mean.round,meta=Schemathesis.delta.round,col sep=comma]{data/apis/branch-coverage-area.csv};
    		\addplot[draw=cyan!80!black,fill=cyan!80!black,opacity=.6] table[x=Api,y=Restler.mean.round,meta=Restler.delta.round,col sep=comma]{data/apis/branch-coverage-area.csv};
    		\legend{\deeprtg\hphantom{S},\rtg\hphantom{S},\aratrl\hphantom{S},Morest\hphantom{S},Schemathesis\hphantom{S},Restler}
    	\end{axis}
    \end{tikzpicture}

    \vspace*{-5pt}

    \begin{tikzpicture}[scale=.5]
    	\begin{axis}[
    		ylabel style={yshift=-0.2cm},
    		ybar=.5pt,
    		axis background/.style={fill=gray!10},
    		bar width=.2cm,
    		width=1.5\textwidth,
    		height=4cm,
    		legend style={at={(.25,.95)},anchor=north,legend columns=-1,draw=none},
    		symbolic x coords={rest-countr,user-mgmt,market,proj-track,feat-serv,ncs,scs,genome-nex,person-ctrl,blog,lang-tool},
    		xtick=data,
    		nodes near coords,
    		nodes near coords align={vertical},
    		ymin=0,ymax=9400,
    		ylabel={Operations (AUC)\vphantom{gplt}},
    		every node near coord/.append style={font=\scriptsize\ttfamily,opacity=1,rotate=90,xshift=7pt,yshift=-5.75pt},
    		point meta=explicit symbolic,
                x tick label style={rotate=35},
    		enlarge x limits=.05,
    		]
    		\addplot[draw=orange!95!black,fill=orange!95!black,opacity=.6] table[x=Api,y=DeepRest.mean.round,col sep=comma]{data/apis/operations-area.csv};
    		\addplot[draw=blue!70,fill=blue!70,opacity=.6] table[x=Api,y=RestTestGen.mean.round,meta=RestTestGen.delta.round,col sep=comma]{data/apis/operations-area.csv};
    		\addplot[draw=olive!60,fill=olive!60,opacity=.6] table[x=Api,y=Arat.mean.round,meta=Arat.delta.round,col sep=comma]{data/apis/operations-area.csv};
    		\addplot[draw=green!70!black,fill=green!70!black,opacity=.6] table[x=Api,y=Morest.mean.round,meta=Morest.delta.round,col sep=comma]{data/apis/operations-area.csv};
    		\addplot[draw=magenta!80!black,fill=magenta!80!black,opacity=.6] table[x=Api,y=Schemathesis.mean.round,meta=Schemathesis.delta.round,col sep=comma]{data/apis/operations-area.csv};
    		\addplot[draw=cyan!80!black,fill=cyan!80!black,opacity=.6] table[x=Api,y=Restler.mean.round,meta=Restler.delta.round,col sep=comma]{data/apis/operations-area.csv};
    	\end{axis}
    \end{tikzpicture}

    \caption{Efficiency results (per API).}\label{fig:efficiency-apis}
\end{figure*}

The results of our evaluation are reported in Figures~\ref{fig:effectiveness-aggregate}, \ref{fig:efficiency-aggregate}, \ref{fig:effectiveness-apis} and \ref{fig:efficiency-apis}. To test for statistical significance of observed differences, we apply the Wilcoxon signed rank test to compare the metrics values achieved by \deeprtg with those by each tool in the comparison, computing the corresponding $p$-values. We assume a significance level of 95\% ($\alpha = 0.05$), that is, we reject the null hypothesis when $p\text{-value} < 0.05$. Values of statistically significant differences are in black in the figures, while non-significant differences are in red.

\begin{figure}[t]
    \begin{tikzpicture}[scale=.5]
    	\begin{axis}[
    		ylabel style={yshift=-0.4cm},
    		ybar=.5pt,
    		axis background/.style={fill=gray!10},
    		legend style={at={(.875,1.35)},anchor=north,legend columns=-1,draw=none},
    		bar width=.3cm,
    		width=.6\textwidth,
    		height=4cm,
    		symbolic x coords={Branch Coverage,Line Coverage,Method Coverage},
    		xtick=data,
    		nodes near coords,
    		nodes near coords align={vertical},
    		ymin=0,ymax=150,
    		ylabel={Area Under Curve\vphantom{gplt}},
    		every node near coord/.append style={font=\scriptsize\ttfamily,opacity=1,rotate=90,xshift=7pt,yshift=-5.75pt},
    		point meta=explicit symbolic,
    		x tick label style={align=center,rotate=0},
    		enlarge x limits=.3,
    		]
    		\addplot[draw=orange!95!black,fill=orange!95!black,opacity=.6] table[x=Metric,y=DeepRest.mean.round,col sep=comma]{data/aggregate/efficiency-1.csv};
    		\addplot[draw=blue!70,fill=blue!70,opacity=.6] table[x=Metric,y=RestTestGen.mean.round,meta=RestTestGen.delta.round,col sep=comma]{data/aggregate/efficiency-1.csv};
    		\addplot[draw=olive!60,fill=olive!60,opacity=.6] table[x=Metric,y=Arat.mean.round,meta=Arat.delta.round,col sep=comma]{data/aggregate/efficiency-1.csv};
    		\addplot[draw=green!70!black,fill=green!70!black,opacity=.6] table[x=Metric,y=Morest.mean.round,meta=Morest.delta.round,col sep=comma]{data/aggregate/efficiency-1.csv};	
    		\addplot[draw=magenta!80!black,fill=magenta!80!black,opacity=.6] table[x=Metric,y=Schemathesis.mean.round,meta=Schemathesis.delta.round,col sep=comma]{data/aggregate/efficiency-1.csv};
    		\addplot[draw=cyan!80!black,fill=cyan!80!black,opacity=.6] table[x=Metric,y=Restler.mean.round,meta=Restler.delta.round,col sep=comma]{data/aggregate/efficiency-1.csv};
    		\legend{\deeprtg\hphantom{S},\rtg\hphantom{S},\aratrl\hphantom{S},Morest\hphantom{S},Schemathesis\hphantom{S},Restler}
    	\end{axis}
    \end{tikzpicture}
    \hspace*{-70pt}
    \begin{tikzpicture}[scale=.5]
    	\begin{axis}[
    		ylabel style={yshift=-0.4cm},
    		ybar=.5pt,
    		axis background/.style={fill=gray!10},
    		bar width=.3cm,
    		width=.4\textwidth,
    		height=4cm,
    		symbolic x coords={Operations,Unique Faults},
    		xtick=data,
    		nodes near coords,
    		nodes near coords align={vertical},
    		ymin=0,ymax=2800,
    		every node near coord/.append style={font=\scriptsize\ttfamily,opacity=1,rotate=90,xshift=7pt,yshift=-5.75pt},
    		point meta=explicit symbolic,
    		x tick label style={align=center,rotate=0},
    		enlarge x limits=.55,
    		]
    		\addplot[draw=orange!95!black,fill=orange!95!black,opacity=.6] table[x=Metric,y=DeepRest.mean.round,col sep=comma]{data/aggregate/efficiency-2.csv};
    		\addplot[draw=blue!70,fill=blue!70,opacity=.6] table[x=Metric,y=RestTestGen.mean.round,meta=RestTestGen.delta.round,col sep=comma]{data/aggregate/efficiency-2.csv};
    		\addplot[draw=olive!60,fill=olive!60,opacity=.6] table[x=Metric,y=Arat.mean.round,meta=Arat.delta.round,col sep=comma]{data/aggregate/efficiency-2.csv};
    		\addplot[draw=green!70!black,fill=green!70!black,opacity=.6] table[x=Metric,y=Morest.mean.round,meta=Morest.delta.round,col sep=comma]{data/aggregate/efficiency-2.csv};	
    		\addplot[draw=magenta!80!black,fill=magenta!80!black,opacity=.6] table[x=Metric,y=Schemathesis.mean.round,meta=Schemathesis.delta.round,col sep=comma]{data/aggregate/efficiency-2.csv};
    		\addplot[draw=cyan!80!black,fill=cyan!80!black,opacity=.6] table[x=Metric,y=Restler.mean.round,meta=Restler.delta.round,col sep=comma]{data/aggregate/efficiency-2.csv};
    	\end{axis}
    \end{tikzpicture}
    \caption{Efficiency results (aggregate).}\label{fig:efficiency-aggregate}
\end{figure}
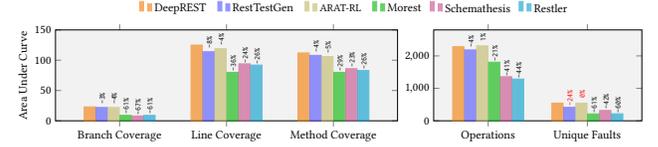

\smallskip
\req{1}~\emph{(Effectiveness).} 
The experimental results illustrating the effectiveness of \deeprtg are presented in Figure~\ref{fig:effectiveness-aggregate} and Figure~\ref{fig:effectiveness-apis}, alongside comparable results from the tools in the comparison for reference. Figure~\ref{fig:effectiveness-aggregate} shows the overall results aggregated by tool, with different bar colors for different tools (\deeprtg is the left-most bar in orange). On the left-hand side of the figure, we report the average branch, line and method coverage among all 11 APIs for each tool. On the right-hand side, we report the average count of successfully tested operations (\twoxx) and faults (\fivexx). 

As we can see from the figure, \deeprtg achieves the highest values for all the metrics among all the considered testing tools, and all the differences are statistically significant according to the Wilcoxon signed rank test (with the only exception of the faults for Restler). The most remarkable difference is for branch coverage that spans from 17\% for \rtg to 77\% for Restler.

Figure~\ref{fig:effectiveness-apis} represents the collected metrics results split by case study. For space reasons, we report the results for two metrics only (results for the other metrics are provided in the replication package~\cite{ReplicationPackage}). The first bar graph represents branch coverage, while the second bar graph the successful operations count. %
This figure confirms the previously observed trend, with \deeprtg outperforming all the other tools in almost all the case study APIs. The very few exceptions include the case of {\em Feature Service} API, for which \aratrl and \rtg achieve an higher score. %

These results suggest that our approach achieved the highest effectiveness because it tested API states that other approaches could not reach. This is probably due to the fact that \deeprtg learned how to assemble more effective sequences of operation calls that could not be assembled just by relying on (producer-consumer) dependencies documented in the API specification.

An interesting case to comment on is {\em Project Tracking System}, where \deeprtg largely overcame other tools (16\% to 62\% higher coverage). The Operation Dependency Graph exported by \rtg for this API contains 62 operations and over 6000 data dependencies. On the one hand, this overwhelming large set of data dependencies probably contains many spurious dependencies, tricking testing tools into building unsuccessful sequences. On the other hand, there are so many dependencies that attempting all the resulting operation orders is quite inefficient. 
The advantage of deep reinforcement learning in this API is that whenever the correct order to test $N$ operations is found, it is learned and reused to try and test a sequence of $N+1$ operations without wasting the testing budget. 

Specifically, in the {\em Project Tracking System} API, only after creating new \inlineyaml{credentials} an \inlineyaml{employee} can be created. Then, a \inlineyaml{project} must be created. Only after successfully taking the API to this state the \inlineyaml{POST /assignment} operation can be tested, resulting in a mandatory sequence of four operations. As a matter of fact, \deeprtg gradually learned this operation ordering, progressively building longer and longer sequences, thus testing more operations on this API.

These results allow us to formulate the following answer to the first research question.

\smallskip
\begin{mdframed}[roundcorner=2pt, backgroundcolor=gray!10]
    \emph{\textbf{Answer to}} \textbf{\req{1}}:~ \deeprtg is the most effective black-box testing tool for REST APIs, demonstrating higher effectiveness than state-of-the-art with respect to branch, line and method coverage, as well as successfully tested operations and revealed unique faults. 
\end{mdframed}

\smallskip
\req{2}~\emph{(Efficiency).} 
The experimental results about the efficiency of \deeprtg are presented in Figure~\ref{fig:efficiency-aggregate} and Figure~\ref{fig:efficiency-apis}.
Figure~\ref{fig:efficiency-aggregate} shows the average results aggregated by tool, with different bar colors for different tools (\deeprtg is the left-most bar in orange). On the left-hand side of the figure, we report the average Area Under Curve (AUC) for branch, line, and method coverage among all 11 APIs for each tool. On the right-hand side, we report the AUC for the successfully tested operations (\twoxx) and the faults (\fivexx) count.

As we can see from the figure, \deeprtg is more efficient in achieving high values for all the metrics than all the other testing tools, with the only exception of \aratrl for the successful operations count. Moreover, almost all the differences are statistically significant according to the Wilcoxon signed rank test (with the only exception of unique faults by \rtg and \aratrl). The most remarkable difference is for the AUC of the successfully tested operations count that spans from 4\% of \rtg to 44\% of Restler, and line coverage that spans between 4\% (\aratrl) to 36\% (Morest).

Figure~\ref{fig:efficiency-apis} shows the results for the AUC of the same metrics, split by case study. For space reasons, we report the results for two metrics only (results for the other metrics are provided in the replication package~\cite{ReplicationPackage}). The first bar graph represents branch coverage, while the second bar graph the successful operations count. \deeprtg appear consistently superior state-of-the-art tools in most of the cases, with few exceptions (already observed for effectiveness); they are \aratrl and \rtg on the branch coverage of {\em Feature Service}. %

Based on these results, we can formulate the following answer to the second research question. 

\smallskip
\begin{mdframed}[roundcorner=2pt, backgroundcolor=gray!10]
    \emph{\textbf{Answer to}} \textbf{\req{2}}:~ \deeprtg demonstrates higher efficiency than state-of-the-art testing tools concerning branch, line and method
    coverage. Regarding successfully tested operations and revealed unique faults, \deeprtg is superior to most state-of-the-art tools except \aratrl.
\end{mdframed}

\subsection{Threats to Validity}
Threats to \emph{internal validity}, impacting empirical results, are due to the metrics adopted to answer the research questions and the configuration of the tools in the comparison. To mitigate these threats, we adopted standard metrics from structural testing (code coverage) and specific metrics for REST API testing (operation coverage~\cite{MartinLopez2019TestCoverage} and fault detection\cite{kim2023adaptive}). Furthermore, the latest versions of the considered testing tools have been used in the comparison, with tools configured as indicated in the corresponding papers. We addressed potential threats due to randomness issues by running each tool 10 times and computing the average results.
Moreover, we applied a statistical test (\ie Wilcoxon signed rank test) that is non-parametric, thus it does not assume experimental data to be normally distributed. 

Threats to \emph{external validity}, impacting the generalization of our findings, are due to the case studies selected for the tools comparison and their limited number. We mitigated these threats by considering the dataset of REST APIs adopted in previous studies~\cite{kim2023adaptive}, in addition to one new API. %

\section{Related Work}

Among automated REST API testing tools, only EvoMaster~\cite{Arcuri2019EvoMaster} adopts a white-box approach. Specifically, test case generation is guided by evolutionary algorithms, whose fitness function is defined in terms of API code coverage and HTTP interactions status code. EvoMaster also provides a black-box version of the tool that only resorts to the API documentation to guide generation.

Literature about black-box REST API testing is broader, comprising various tools implementing different testing strategies. We already mentioned approaches, such as \rler~\cite{Atlidakis2019RESTler} and \rtg~\cite{Corradini2022}, that use data dependencies to prioritize operations testing. \rler generates sequences of HTTP interactions by exploiting producer-consumer dependencies contained in the OAS, targeting internal server failures. Instead, \rtg exploits the Operation Dependency Graph, embedding data dependencies between operations, to craft meaningful test cases in nominal and error scenarios. As already pointed out, such approaches might be biased by spurious data dependencies or miss implicit operation dependencies that are instead exploited by \deeprtg.%

From a general viewpoint, QuickREST~\cite{Karlsson2020QuickREST} performs property-based testing of REST APIs, generating test cases with the aim of verifying whether an API complies with some properties documented in its OAS. Similarly, Schemathesis~\cite{schemathesis} detects faults by checking response compliance in OpenAPI or GraphQL APIs via property-based testing. Morest~\cite{morest} exploits a Property Graph, dynamically updated during testing, to model the behavior of the API under test. Knowledge from the graph is used to craft meaningful test sequences. RESTest~\cite{MartinLopez2020RESTest} is a tool that provides inter-parameter dependencies testing, producing nominal and faulty test cases. RestCT~\cite{restct} leverages combinatorial testing to generate test cases for REST APIs based on the OAS. Dredd~\cite{dredd} is a tool testing REST APIs by comparing actual responses with expected ones, checking their status code, header, and body. Tcases~\cite{tcases} is a model-based tool, leveraging the OAS to systematically build an input space model. Subsequently, it generates test cases covering valid input dimensions and checking response status codes for validation.

REST API fuzzers~\cite{APIFuzzer,FuzzLightyear,FuzzySwagger,SwaggerFuzzer,TnTFuzzer} are black-box tools that generate new test cases starting from previously recorded HTTP traffic: they fuzz and replay new HTTP requests in order to find faults. Some of them~\cite{FuzzLightyear,FuzzySwagger,SwaggerFuzzer} also exploit OASs.

Reinforcement learning has been recently adopted in software testing, focusing primarily on web and mobile applications. In particular, Zheng et al.~\cite{Zheng-ICSE-2021} and Pan et al.~\cite{Pan-ISSTA-2020} propose automatic testing approaches based on curiosity-driven reinforcement learning for web clients and Android apps, respectively. Vuong and Takada~\cite{Vong-Atest-2018} and Koroglu et al.~\cite{Koroglu-ICST-2018} also apply reinforcement learning to automated testing of Android apps, the latter adopting an exploration method based on Q-Learning. Adamo et al.~\cite{Adamo-Atest-2018} present a reinforcement learning-based technique specifically designed for Android GUI testing, while Koroglu and Sen~\cite{Koroglu-STVR-2021} present a reinforcement learning-based method for generating functional tests from UI test scenarios for Android apps. Mariani et al.~\cite{Mariani-ICST-2012} proposed AutoBlackTest, an automatic black-box testing approach for interactive applications.

The closest work is \aratrl~\cite{kim2023adaptive}, which exploits reinforcement learning to test REST APIs. Utilizing Q-Learning, \aratrl determines the priority of operations to test initially by considering the frequency of parameters in the API specification (that can be seen as a sort of data dependencies-based initialization). It subsequently refines this prioritization based on the HTTP interactions with the API. Nevertheless, such refinement is limited by the learning strategy adopted by \aratrl: agents always receive a high penalty when successfully tested operations are considered again. This promotes exploration of not yet tested operations only, discouraging agents from crafting complex sequences of operations. Indeed, to spot implicit dependencies, considering already successfully tested operations is crucial since they may trigger operations that otherwise are unlikely to be exercised. 
In \deeprtg, instead, deep reinforcement learning starts with an empty experience, avoiding any possible bias from data dependencies. Moreover, \deeprtg explicitly models (an approximation of) the internal state of the REST API under test, and agents are encouraged to test operations multiple times in different API states. This may yield corner case preconditions triggering hard-to-test operations, corresponding to implicit dependencies that can be spotted by crafting complex sequences of operation invocations only. This is possible due to the \deeprtg API state representation encoding the history of past successes, %
which is used as a guide for operation sequencing. This complex state representation can not be modeled by Q-Learning and requires a \emph{deep} reinforcement learning approach.

\section{Conclusion}
Black-box REST API testing tools ground their test case generation strategies on the information contained in the OAS of the API under test. This poses limitations on 
those APIs that contain implicit operation dependencies and value conditions. Such constraints cannot be retrieved from the OAS, making testing tools blind with respect to potentially crucial parts of the API business logic.

In this paper, we showed how such hidden API constraints can be learned from API interactions attempted at testing time, even in a black-box setting. We indeed proposed the first REST API testing approach based on deep reinforcement learning, having the twofold objective of computing an effective ordering of API operations to test, encompassing (explicit and implicit) operation dependencies, and selecting accurate input data for operation parameters. Operations ordering inference is guided by \emph{exploration} of the API under test, while input data generation leverages \emph{experience} gained on successful API interactions. Finally, \deeprtg \emph{intensifies} testing by mutating successful API interactions in order to achieve higher test coverage and collaterally increasing experience.  

Empirical evidence showed that the proposed approach %
results in boosting the performance of REST API testing. Indeed, \deeprtg is shown to overcome state-of-the-art approaches, both in terms of effectiveness and efficiency. %

\section*{Data Availability}
All the material needed to replicate our experiments is available on Zenodo~\cite{ReplicationPackage}.

\bibliographystyle{IEEEtran}
\bibliography{bib}

\begin{thebibliography}{10}
\providecommand{\url}[1]{#1}
\csname url@samestyle\endcsname
\providecommand{\newblock}{\relax}
\providecommand{\bibinfo}[2]{#2}
\providecommand{\BIBentrySTDinterwordspacing}{\spaceskip=0pt\relax}
\providecommand{\BIBentryALTinterwordstretchfactor}{4}
\providecommand{\BIBentryALTinterwordspacing}{\spaceskip=\fontdimen2\font plus
\BIBentryALTinterwordstretchfactor\fontdimen3\font minus \fontdimen4\font\relax}
\providecommand{\BIBforeignlanguage}[2]{{%
\expandafter\ifx\csname l@#1\endcsname\relax
\typeout{** WARNING: IEEEtran.bst: No hyphenation pattern has been}%
\typeout{** loaded for the language `#1'. Using the pattern for}%
\typeout{** the default language instead.}%
\else
\language=\csname l@#1\endcsname
\fi
#2}}
\providecommand{\BIBdecl}{\relax}
\BIBdecl

\bibitem{Arcuri-TOSEM23-Survey}
A.~Golmohammadi, M.~Zhang, and A.~Arcuri, ``Testing restful apis: A survey,'' \emph{ACM Trans. Softw. Eng. Methodol.}, vol.~33, no.~1, nov 2023.

\bibitem{Atlidakis2019RESTler}
\BIBentryALTinterwordspacing
V.~Atlidakis, P.~Godefroid, and M.~Polishchuk, ``{RESTler}: {Stateful} {REST} {API} fuzzing,'' in \emph{Proceedings of the 41st International Conference on Software Engineering}, ser. {ICSE} '19.\hskip 1em plus 0.5em minus 0.4em\relax Piscataway, NJ, USA: IEEE Press, 2019, pp. 748--758. [Online]. Available: \url{https://doi.org/10.1109/ICSE.2019.00083}
\BIBentrySTDinterwordspacing

\bibitem{Corradini2022}
\BIBentryALTinterwordspacing
D.~Corradini, A.~Zampieri, M.~Pasqua, E.~Viglianisi, M.~Dallago, and M.~Ceccato, ``Automated black-box testing of nominal and error scenarios in restful apis,'' \emph{Software Testing, Verification and Reliability}, Jan. 2022. [Online]. Available: \url{https://doi.org/10.1002/stvr.1808}
\BIBentrySTDinterwordspacing

\bibitem{morest}
Y.~Liu, Y.~Li, G.~Deng, Y.~Liu, R.~Wan, R.~Wu, D.~Ji, S.~Xu, and M.~Bao, ``Morest: Model-based restful api testing with execution feedback,'' in \emph{Proceedings of the 44th International Conference on Software Engineering}.\hskip 1em plus 0.5em minus 0.4em\relax New York, NY, USA: ACM, 2022, pp. 1406--1417.

\bibitem{Atlidakis2020SecurityProperties}
\BIBentryALTinterwordspacing
V.~Atlidakis, P.~Godefroid, and M.~Polishchuk, ``Checking security properties of cloud service {REST} {APIs},'' in \emph{13th {IEEE} International Conference on Software Testing, Validation and Verification, {ICST} 2020, Porto, Portugal, October 24-28, 2020}.\hskip 1em plus 0.5em minus 0.4em\relax {IEEE}, 2020, pp. 387--397. [Online]. Available: \url{https://doi.org/10.1109/ICST46399.2020.00046}
\BIBentrySTDinterwordspacing

\bibitem{MartinLopez2021DependenciesAPI}
A.~{Martin-Lopez}, S.~{Segura}, C.~{Muller}, and A.~{Ruiz-Cortes}, ``Specification and automated analysis of inter-parameter dependencies in web {APIs},'' \emph{{IEEE} Transactions on Services Computing}, pp. 1--1, 2021.

\bibitem{kim2023adaptive}
M.~Kim, S.~Sinha, and A.~Orso, ``Adaptive rest api testing with reinforcement learning,'' in \emph{IEEE/ACM International Conference on Automated Software Engineering}, 2023.

\bibitem{ReplicationPackage}
``Replication package,'' \url{https://zenodo.org/records/11525389}.

\bibitem{Fielding2000ArchitecturalStyle}
R.~T. Fielding, \emph{Architectural styles and the design of network-based software architectures}.\hskip 1em plus 0.5em minus 0.4em\relax University of California, Irvine Doctoral dissertation, 2000, vol.~7.

\bibitem{ppo}
J.~Schulman, F.~Wolski, P.~Dhariwal, A.~Radford, and O.~Klimov, ``Proximal policy optimization algorithms,'' \emph{CoRR}, vol. abs/1707.06347, 2017.

\bibitem{Watkins1992}
C.~J. C.~H. Watkins and P.~Dayan, ``Q-learning,'' \emph{Machine Learning}, vol.~8, no.~3, pp. 279--292, may 1992.

\bibitem{Zheng-ICSE-2021}
Y.~Zheng, Y.~Liu, X.~Xie, Y.~Liu, L.~Ma, J.~Hao, and Y.~Liu, ``Automatic web testing using curiosity-driven reinforcement learning,'' in \emph{2021 IEEE/ACM 43rd International Conference on Software Engineering (ICSE)}, 2021, pp. 423--435.

\bibitem{Pan-ISSTA-2020}
M.~Pan, A.~Huang, G.~Wang, T.~Zhang, and X.~Li, ``Reinforcement learning based curiosity-driven testing of android applications,'' in \emph{Proceedings of the 29th ACM SIGSOFT International Symposium on Software Testing and Analysis}.\hskip 1em plus 0.5em minus 0.4em\relax New York, NY, USA: ACM, 2020, pp. 153--164.

\bibitem{openai}
{OpenAI}, ``Proximal policy optimization,'' 2023, \url{https://openai.com/research/openai-baselines-ppo}.

\bibitem{raffin2021stable}
A.~Raffin, A.~Hill, A.~Gleave, A.~Kanervisto, M.~Ernestus, and N.~Dormann, ``Stable-baselines3: Reliable reinforcement learning implementations,'' \emph{Journal of Machine Learning Research}, vol.~22, no. 268, pp. 1--8, 2021.

\bibitem{GptForAll}
{nomic-ai}, ``Gpt4all,'' 2024, \url{https://gpt4all.io}.

\bibitem{kim2023enhancing}
M.~Kim, D.~Corradini, S.~Sinha, A.~Orso, M.~Pasqua, R.~Tzoref-Brill, and M.~Ceccato, ``Enhancing rest api testing with nlp techniques,'' in \emph{Proceedings of the 32nd ACM SIGSOFT International Symposium on Software Testing and Analysis}, 2023, pp. 1232--1243.

\bibitem{blogApi}
{GitHub}, ``Blog rest api,'' 2024, \url{https://github.com/osopromadze/Spring-Boot-Blog-REST-API}.

\bibitem{schemathesis}
Z.~Hatfield-Dodds and D.~Dygalo, ``Deriving semantics-aware fuzzers from web api schemas,'' in \emph{Proceedings of the ACM/IEEE 44th International Conference on Software Engineering: Companion Proceedings}.\hskip 1em plus 0.5em minus 0.4em\relax New York, NY, USA: ACM, 2022, pp. 345--346.

\bibitem{MartinLopez2019TestCoverage}
\BIBentryALTinterwordspacing
A.~Martin-Lopez, S.~Segura, and A.~Ruiz-Cort\'{e}s, ``Test coverage criteria for restful web apis,'' in \emph{Proceedings of the 10th {ACM} {SIGSOFT} International Workshop on Automating {TEST} Case Design, Selection, and Evaluation}, ser. A-TEST 2019.\hskip 1em plus 0.5em minus 0.4em\relax New York, NY, USA: Association for Computing Machinery, 2019, pp. 15--21. [Online]. Available: \url{https://doi.org/10.1145/3340433.3342822}
\BIBentrySTDinterwordspacing

\bibitem{corradini2021restats}
D.~Corradini, A.~Zampieri, M.~Pasqua, and M.~Ceccato, ``Restats: A test coverage tool for restful apis,'' in \emph{2021 IEEE International Conference on Software Maintenance and Evolution (ICSME)}.\hskip 1em plus 0.5em minus 0.4em\relax IEEE, 2021, pp. 594--598.

\bibitem{jacoco}
{Jacoco}, ``{JaCoCo Java Code Coverage Library},'' 2023, \url{https://github.com/jacoco/jacoco}.

\bibitem{Arcuri2019EvoMaster}
A.~Arcuri, ``{RESTful} {API} automated test case generation with {E}vomaster,'' \emph{{ACM} Transactions on Software Engineering and Methodology ({TOSEM})}, vol.~28, no.~1, p.~3, 2019.

\bibitem{Karlsson2020QuickREST}
S.~{Karlsson}, A.~{Causevic}, and D.~{Sundmark}, ``{QuickREST}: Property-based test generation of {OpenAPI}-described {RESTful} {APIs},'' in \emph{2020 {IEEE} 13th International Conference on Software Testing, Validation and Verification ({ICST})}, 2020, pp. 131--141.

\bibitem{MartinLopez2020RESTest}
\BIBentryALTinterwordspacing
A.~Martin{-}Lopez, S.~Segura, and A.~Ruiz{-}Cort{\'{e}}s, ``{RESTest}: Black-box constraint-based testing of {RESTful} web {APIs},'' in \emph{Service-Oriented Computing - 18th International Conference, {ICSOC} 2020, Dubai, United Arab Emirates, December 14-17, 2020, Proceedings}, ser. Lecture Notes in Computer Science, E.~Kafeza, B.~Benatallah, F.~Martinelli, H.~Hacid, A.~Bouguettaya, and H.~Motahari, Eds., vol. 12571.\hskip 1em plus 0.5em minus 0.4em\relax Springer, 2020, pp. 459--475. [Online]. Available: \url{https://doi.org/10.1007/978-3-030-65310-1\_33}
\BIBentrySTDinterwordspacing

\bibitem{restct}
H.~Wu, L.~Xu, X.~Niu, and C.~Nie, ``Combinatorial testing of restful apis,'' in \emph{Proceedings of the 44th International Conference on Software Engineering}.\hskip 1em plus 0.5em minus 0.4em\relax New York, NY, USA: ACM, 2022, pp. 426--437.

\bibitem{dredd}
{apiaryio}. (2023) {Dredd}. \url{https://github.com/apiaryio/dredd}.

\bibitem{tcases}
{Cornutum}. (2023) {Tcases}. \url{https://github.com/Cornutum/tcases}.

\bibitem{APIFuzzer}
{API Fuzzer}. (2022) {API Fuzzer}. \url{https://github.com/KissPeter/APIFuzzer}.

\bibitem{FuzzLightyear}
{Fuzz-Lightyear}, ``{Fuzz-Lightyear},'' 2022, \url{https://github.com/Yelp/fuzz-lightyear}.

\bibitem{FuzzySwagger}
{Fuzzy-Swagger}, ``{Fuzzy-Swagger},'' 2022, \url{https://github.com/namuan/fuzzy-swagger}.

\bibitem{SwaggerFuzzer}
{Swagger-Fuzzer}, ``{Swagger-Fuzzer},'' 2022, \url{https://github.com/Lothiraldan/swagger-fuzzer}.

\bibitem{TnTFuzzer}
{TnT-Fuzzer}, ``{TnT-Fuzzer},'' 2022, \url{https://github.com/Teebytes/TnT-Fuzzer}.

\bibitem{Vong-Atest-2018}
T.~A.~T. Vuong and S.~Takada, ``A reinforcement learning based approach to automated testing of android applications,'' in \emph{Proceedings of the 9th ACM SIGSOFT International Workshop on Automating TEST Case Design, Selection, and Evaluation}.\hskip 1em plus 0.5em minus 0.4em\relax New York, NY, USA: ACM, 2018, pp. 31--37.

\bibitem{Koroglu-ICST-2018}
Y.~Koroglu, A.~Sen, O.~Muslu, Y.~Mete, C.~Ulker, T.~Tanriverdi, and Y.~Donmez, ``{QBE}: Qlearning-based exploration of android applications,'' in \emph{2018 IEEE 11th International Conference on Software Testing, Verification and Validation (ICST)}, 2018, pp. 105--115.

\bibitem{Adamo-Atest-2018}
D.~Adamo, M.~K. Khan, S.~Koppula, and R.~Bryce, ``Reinforcement learning for android gui testing,'' in \emph{Proceedings of the 9th ACM SIGSOFT International Workshop on Automating TEST Case Design, Selection, and Evaluation}.\hskip 1em plus 0.5em minus 0.4em\relax New York, NY, USA: ACM, 2018, pp. 2--8.

\bibitem{Koroglu-STVR-2021}
Y.~Koroglu and A.~Sen, ``Functional test generation from ui test scenarios using reinforcement learning for android applications,'' \emph{Software Testing, Verification and Reliability}, vol.~31, no.~3, p. e1752, 2021.

\bibitem{Mariani-ICST-2012}
L.~Mariani, M.~Pezze, O.~Riganelli, and M.~Santoro, ``{AutoBlackTest}: Automatic black-box testing of interactive applications,'' in \emph{2012 IEEE Fifth International Conference on Software Testing, Verification and Validation}, 2012, pp. 81--90.

\end{thebibliography}

\end{document}